\documentclass{moriond}
\usepackage{xspace}
\usepackage{braket}
\usepackage{enumitem}
\newcommand{\Py}{\textsc{Pythia}\xspace}
\newcommand{\dndeta}{\ensuremath{\langle dN_\mathrm{ch}/d\eta \rangle_{|\eta|<0.5}}\xspace}

\def\be{\begin{equation}}
\def\ee{\end{equation}}
\def\bea{\begin{eqnarray}}
\def\eea{\end{eqnarray}}

\newcommand{\Photo}{}

\begin{document}
\vspace*{4cm}
\title{Strangeness enhancement in pp collisions from string closepacking in \Py 8.3}

\author{Javira Altmann, Peter Skands}

\address{School of Physics and Astronomy, Monash University, VIC-3800, Australia}

\author{Lorenzo Bernardinis$^{\ast}$\let\thefootnote\relax\footnotetext{\hspace{-0.5em}$^{\ast}$speaker}, Valentina Zaccolo}
\address{Dipartimento di Fisica dell’Università and Sezione INFN, Via Alfonso Valerio 2, Trieste, 34127,
Italy}
\maketitle\abstracts{
Measurements at LHC show an increased production of strange hadrons with charged multiplicity in pp collisions, which is not described by the Lund String Model (with the Monash tune) implemented in \Py. 
This work investigates \emph{string closepacking}, a mechanism invoked during hadronization where overlapping strings create a background field that increases the effective string tension. This reduces strangeness suppression, effectively enhancing production. The model also incorporates an option for ``popcorn destructive interference'', which suppresses baryon production, to address the non-strange $p/\pi$ ratio, utilizing color algebra arguments; and an option for ``strange junctions'', which enhances strangeness specifically within the baryon sector.
The Trieste tunes of this model to LHC data are presented. The closepacking model is in qualitative agreement with many of the salient particle ratios, although the $\Xi_c/D$ ratio and the shape of $p_\perp$ spectra remain challenging to account for.
Overall, the closepacking model with the Trieste tunes provides a competitive description of enhanced strangeness production in pp collisions, improving upon existing \Py models while avoiding excessive proton yields.}

\section{Hadronization in \Py}
The Lund String Model~\cite{Andersson:1983jt,Sjostrand:1984ic} (LSM) is a semi-classical hadronization model implemented in \Py~\cite{PYTHIA_8_3_manual} Monte Carlo (MC) event generator, which is based on QCD confinement and relativistic string dynamics. At non perturbative scales, the $q\bar{q}$ confining potential is approximated as a linear field $\kappa r$, where $r$ is their relative distance. In this framework, the confining field is interpreted as a Lund string: a 1+1 dimensional object with a constant string tension $\kappa\sim 1$ GeV/fm that forms between color-connected charges.

During the hadronization process color-connected charges stretch the color field, forming a string between them. Given large confining potential, it can become energetically favorable to break the string via quantum tunneling, spontaneously producing new $q\bar{q}$ (or diquark-antidiquark) pairs from the vacuum (Lund fragmentation). This iterative process continues until the available energy is sufficient for further breaks, ultimately resulting in the observed final-state hadrons. 
The LSM models (di)quark production via a QCD equivalent of the QED Schwinger mechanism~\footnote{$e^+e^-$ spontaneous pair creation in the presence of a strong electric field~\cite{SchwingerQED}.}, with a tunneling probability:
\begin{equation}
    \label{eqn:schwinger}
    \exp{\biggl(-\frac{\pi m_{Tq}^2}{\kappa}\biggr)} =  \exp{\biggl(-\frac{\pi m_{q}^2}{\kappa}\biggr)} \exp{\biggl(-\frac{\pi p_{Tq}^2}{\kappa}\biggr)},
\end{equation}
where $m_{Tq} = \sqrt{m_q^2 + p_{Tq}^2}$ is the quark transverse mass and $m_q$ is the effective quark mass. The first term in Eq.~\ref{eqn:schwinger} governs flavor suppression. Due to their large masses, the production of charm and bottom quarks during the soft hadronization process is negligible, while the production probability of strange quarks ($s$) relative to up and down ($u,d$) is governed by $P(s:u,d) = \exp{[-\pi (m_{s}^2-m_{u,d}^2)/\kappa}]$. The diquark-to-quark production probability $P(qq:q)$, which regulates the baryon-to-meson ratio, also scales with the string tension $\kappa$, albeit following a different functional form~\cite{Altmann:2025afh}. Finally, the second term in Eq.~\ref{eqn:schwinger} implies a gaussian transverse momentum $p_T$ spectrum of the string breaks.

Despite these probabilities being tuned to LEP data, the $p/\pi$ ratio as a function of multiplicity, \dndeta, measured by ALICE in pp collisions~\cite{ALICE_pp_2017}, is overestimated by \Py using the default Monash tune~\cite{Skands:2014pea}, which implements LSM. This discrepancy suggests that the $p/\pi$ ratio is lower at LHC than at LEP, highlighting a lack in modeling baryon production in denser string environments. ALICE also measured $p_T$-integrated yield ratios of strange hadrons to charged pions, revealing an enhancement of strangeness production as a function of multiplicity~\cite{ALICE_pp_2017}. The Monash tune~\cite{Skands:2014pea} yields to a flat trend, significantly underestimating data at higher multiplicities~\cite{ALICE_pp_2017}, as also shown in Fig.~\ref{fig:ratios} for the $\Lambda/\pi$ and $\Xi/\pi$ ratios.

These observations motivated the \textit{Rope Hadronization} model~\cite{Bierlich:2014xba}, which posits that in high string density environments, such as pp collisions at LHC, overlapping strings act collectively rather than independently. Their overlap results in an enhanced effective string tension $\tilde{\kappa}$, which modulates flavor production probabilities. Specifically, the strangeness suppression factor becomes $\tilde{P}(s:u,d)=P(s:u,d)^{\kappa/\tilde{\kappa}}$; thus an increased $\tilde{\kappa}$ reduces the suppression, leading to the observed strangeness enhancement. While the Rope model successfully reproduces the trend of strange hadron ratios measured by ALICE~\cite{ALICE_pp_2017}, it overestimates the $p/\pi$ ratio by more than $40\%$ (see Fig.~\ref{fig:ratios}).

\section{Closepacking}
\label{sec:closepacking}
Closepacking~\cite{Altmann:2025afh} is a novel hadronization model for \Py, also designed to reproduce the strange-hadron enhancement measured by ALICE~\cite{ALICE_pp_2017} without overestimating $p/\pi$.
Similar to Rope Hadronization, it assumes that string overlaps increase the effective tension $\tilde{\kappa}$. However, Closepacking is fully implemented in momentum space as the LSM, thereby avoiding the computational cost of the space-time modeling of string breaks required by the Rope framework.

The model adopts Casimir scaling from LQCD~\cite{Bali:2000un}, where the total color charge of an irreducible $SU(3)$ representation $R$, with indices $(p,q)$, is given by the quadratic Casimir operator $C_2(p,q)$~\cite{Altmann:2025afh}. Assuming that the color reconnection step has screened the system into its smallest irreducible equivalents, the effective string tension $\tilde{\kappa}$ scales with the colour factor $C_R=C_2(p,q)$ of the multiplet $R$ relative to $C_F=4/3$, which characterizes the fundamental $q\bar{q}$ string:
\begin{equation}
    \label{eqn:tension_full}
    \tilde{\kappa} = \frac{C_R}{(p+q)C_F}\kappa=\biggl[1+c_p \biggl(\frac{p+\omega q}{1+p_{T,\mathrm{had}}^2/p_{T0}^2}\biggr)\biggr] \kappa.
\end{equation}
Here, $p$ and $q$ represent the number of parallel and antiparallel strings with respect to the color flux for the representation $R$. The modified tension in Closepacking~\cite{Altmann:2025afh} is parametrized with the tunable parameters $c_p$ and $\omega$, which account for deviations from exact Casimir scaling, valid only for pointlike static color charges. Moreover, to account for the fact that high-$p_T$ partons fragment far from the beam axis and experience reduced closepacking effects compared to partons near the beam, $\tilde{\kappa}$ is also modulated by a $p_{T}$-dependent term, where $p_{T0}$ is a regularization parameter and $p_{T,\mathrm{had}}$ is the transverse momentum of the produced hadron.

\subsection{Popcorn destructive interference}
Baryon production in \Py relies primarily on two mechanisms: diquark-antidiquark pair production via string breaks~\cite{PYTHIA_8_3_manual} using the Schwinger mechanism, and junction string topologies~\cite{Christiansen:2015yqa}. The diquark picture can be generalized by the \textit{popcorn mechanism}, where diquarks form via two successive color fluctuations on a string, rather than direct tunneling from the vacuum. In high-density environments, the presence of a nearby string can interfere with the color fluctuations on the other string. Specifically, the fluctuation can interact with and break the nearby string, thereby preventing the diquark formation from completing and consequently reducing the overall baryon production rate. This \textit{popcorn destructive interference} mechanism~\cite{Altmann:2025afh}, introduced alongside closepacking, scales the diquark-to-quark production probability $P(qq:q)$ and can resolve the $p/\pi$ overestimation. 

\subsection{Strange Junctions}
The simultaneous overprediction of $p/\pi$ and underprediction of multi-strange baryons suggests that strangeness enhancement may be localized within the baryon sector.
The color field near a junction, namely a Y-shape string topology with three string legs connected at one end, due to the vicinity of the three legs, could not be approximated with a linear potential as in a dipole string with $q$ and $\bar{q}$ as endpoints. An enhanced energy density is expected in the field near a junction, leading to a higher string tension that affects string breaks in its vicinity and increases strangeness production. This mechanism, called \textit{strange junctions} and introduced in this work~\cite{Altmann:2025afh}, modulates the strangeness probability for breaks next to the junctions with the parameter $J_s\in[0,1]$:
\begin{equation}
    P^{'}(s:u,d) = P(s:u,d)^{1-J_s}.
\end{equation}
$J_s = 1$ implies maximal strangeness around a junction, while $J_s = 0$ recovers the baseline probability. 

\section{Results: Closepacking Trieste Tunes}
A first tuning of the Closepacking model was performed in this work~\cite{Altmann:2025afh} with the aim of reproducing strange hadron-to-pion ratios, as well as the $p/\pi$ ratio, measured by ALICE~\cite{ALICE_pp_2017}. The tuning also ensures consistency with other pp observables, such as $\braket{p_T}$ vs $N_\mathrm{ch}$ and $dN_\mathrm{ch}/d\eta$ distributions.

\begin{figure}[ht]
    \begin{minipage}{0.49\linewidth}
        \centerline{\includegraphics[width=0.95\linewidth]{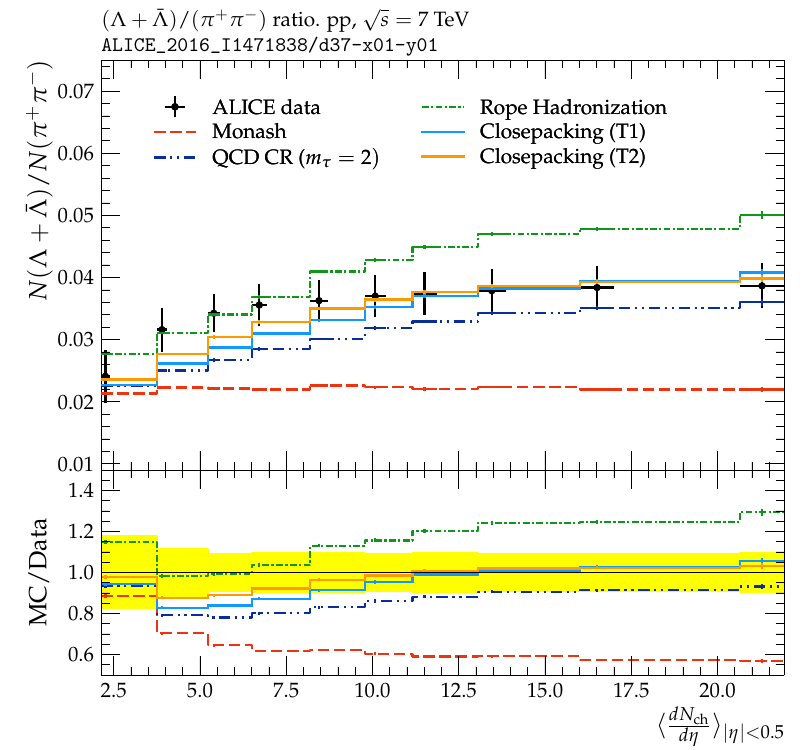}}
    \end{minipage}
    \hfill
    \begin{minipage}{0.49\linewidth}
        \centerline{\includegraphics[width=0.95\linewidth]{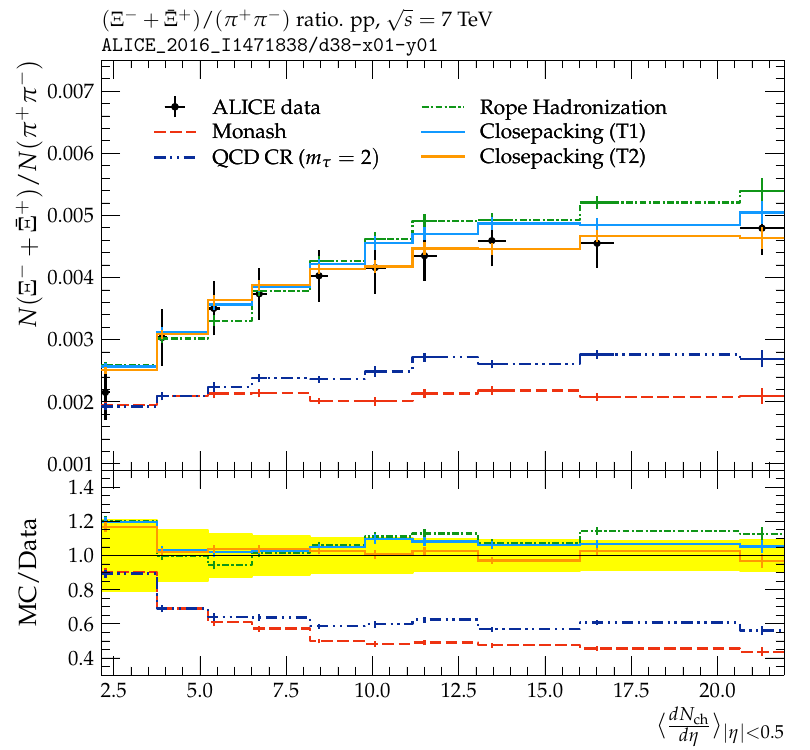}}
    \end{minipage}
    \\
    \begin{minipage}{0.49\linewidth}
        \centerline{\includegraphics[width=0.95\linewidth]{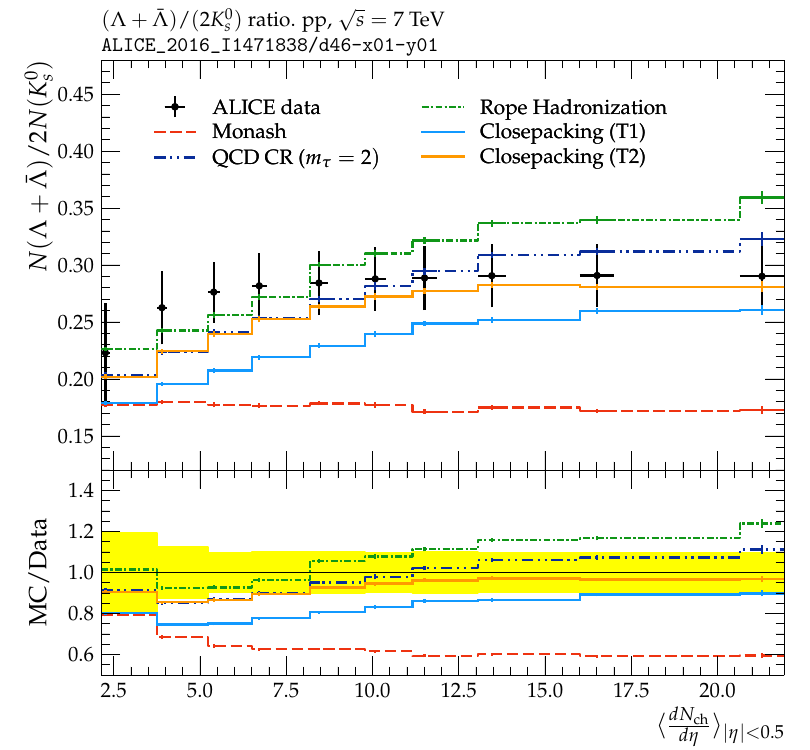}}
    \end{minipage}
    \hfill
    \begin{minipage}{0.49\linewidth}
        \centerline{\includegraphics[width=0.95\linewidth]{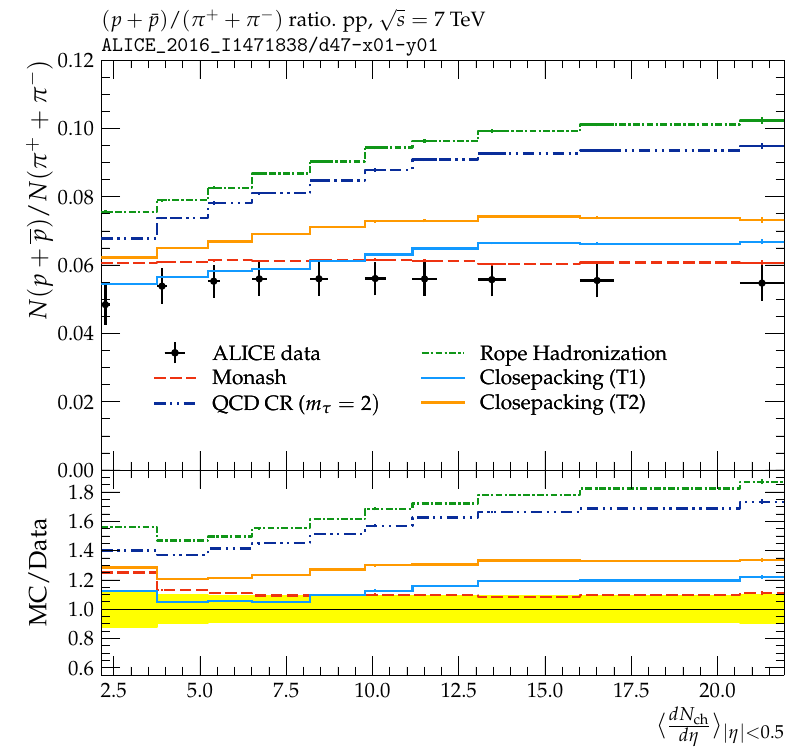}}
    \end{minipage} 
    \caption[]{\Py predictions of different models for $(\Lambda+\overline{\Lambda})/(\pi^+ + \pi^-)$ (top left), $(\Xi^-+\overline{\Xi}^+)/(\pi^+ + \pi^-)$ (top right), $(\Lambda+\overline{\Lambda})/(2K_\mathrm{S}^0)$ (bottom left), $(p+\overline{p})/(\pi^+ + \pi^-)$ (bottom right) ratios as a function of \dndeta, compared against ALICE data~\cite{ALICE_pp_2017}.}
    \label{fig:ratios}
\end{figure}

The model parameters were tuned using the Professor 2 framework~\cite{Buckley:2009bj}. This tool parametrizes the Monte Carlo response using a continuous polynomial interpolation $f_b(\vec{p})$ for each observable bin $b$, constructed from an initial sampling of the $P$-dimensional parameter space. The optimal parameter set $\vec{p}_\mathrm{best}$ is subsequently identified by minimizing a weighted $\chi^2$ function. Professor works along with the Rivet 3 framework~\cite{Bierlich:2019rhm}, which facilitates generator-independent data comparisons and provides the plotting utilities used as baseline for Fig.~\ref{fig:ratios}.

The results of this work are two Trieste tunes: Closepacking T1 and Closepacking T2. They both implement closepacking, popcorn destructive interference and strange junctions. In the T1 tune, a higher strength of destructive interference with respect to the T2 tune results in a lower $p/\pi$ ratio, comparable to the Monash tune~\cite{Skands:2014pea} prediction, whereas the T2 tune overestimates the data by approximately $20\%$ (see Fig.~\ref{fig:ratios}).

Figure~\ref{fig:ratios} compares the predictions of the Trieste Tunes for the $\Lambda/\pi$, $\Xi/\pi$, $\Lambda/K_\mathrm{S}^0$ and $p/\pi$ ratios as a function of \dndeta against ALICE data~\cite{ALICE_pp_2017} and other \Py configurations. These observables also served as tuning inputs. Predictions were generated using PYTHIA 8.316 with the following configurations:
\begin{enumerate}[noitemsep]
    \item Monash: LSM model with the Monash tune~\cite{Skands:2014pea};
    \item QCD CR ($m_\tau=2$): QCD-based color reconnection tune with time-dilation mode 2~\cite{Christiansen:2015yqa};
    \item Ropes default: Rope Hadronization model as used in the Closepacking paper~\cite{Altmann:2025afh};
    \item Closepacking (T1): Closepacking Trieste 1 tune~\cite{Altmann:2025afh};
    \item Closepacking (T2): Closepacking Trieste 2 tune~\cite{Altmann:2025afh};
\end{enumerate}
Unlike the flat Monash baseline, Closepacking successfully reproduces the multiplicity-dependent strangeness enhancement, improving upon the Rope predictions for $\Lambda/\pi$ and $p/\pi$, and outperforming the QCD CR ($m_\tau=2$). Some tension still persists since T2 overestimates $p/\pi$ by $\sim20\%$, while T1 underestimates the $\Lambda/K_\mathrm{S}^0$ ratio by a similar margin. Nonetheless, both the Trieste tunes prove to be competitive with, and in several cases superior to, previous modeling attempts.

\section*{References}
\bibliography{Bernardinis}


\end{document}